\documentclass[epj,nopacs]{svjour}
\usepackage{graphics}
\usepackage{xcolor}

\usepackage{latexsym}   
\usepackage{enumerate}
\usepackage{rotating,booktabs,multirow}
\begin{document}
\title{Understanding the energy dependence of $B_2$ in heavy ion collisions: Interplay of volume and space-momentum correlations}
\author{Vincent Gaebel$^{1}$, Michel Bonne$^{1}$, Tom Reichert$^{1}$, Ajdin Burnic$^{5}$, Paula~Hillmann$^{1,2,3,4}$, Marcus~Bleicher$^{1,2,3,4}$}

%
\institute{Institut f\"ur Theoretische Physik, Goethe Universit\"at Frankfurt, Max-von-Laue-Strasse 1, 60438 Frankfurt am Main, Germany 
\and GSI Helmholtzzentrum f\"ur Schwerionenforschung GmbH, Planckstr. 1, 64291 Darmstadt, Germany 
\and John von Neumann-Institut f\"ur Computing, Forschungzentrum J\"ulich,
52425 J\"ulich, Germany 
\and Helmholtz Research Academy Hesse for FAIR, Campus Frankfurt, Germany
\and University of Sarajevo, Sarajevo, Bosnia and Herzegovina}
\date{Received: date / Revised version: date}
%
\abstract{
The deuteron coalescence parameter $B_2$ in proton+proton and nucleus+nucleus collisions in the energy range of $\sqrt{s_{NN}}=$ 900 - 7000 GeV for proton+proton and $\sqrt{s_{NN}}=$ 2 - 2760 GeV for nucleus+nucleus collisions is analyzed with the Ultrarelativistic Quantum Molecular Dynamics (UrQMD) transport model, supplemented by an event-by-event phase space coalescence model for deuteron and anti-deuteron production. The results are compared to data by the E866, E877, PHENIX, STAR and ALICE experiments. The $B_2$ values are calculated from the final spectra of protons and deuterons. At lower energies, $\sqrt{s_{NN}}\leq 20$ GeV, $B_2$ drops drastically with increasing energy. The calculations confirm that this is due to the increasing freeze-out volume reflected in $B_2\sim 1/V$. At higher energies, $\sqrt{s_{NN}}\geq 20$ GeV, $B_2$ saturates at a constant level. This qualitative change and the vanishing of the volume suppression is shown to be due to the development of strong radial flow with increasing energy. The flow leads to strong space-momentum correlations which counteract the volume effect.}
%
%
\titlerunning{Analysis of the deuteron coalescence parameter $B_{2}$ with the UrQMD model}
\authorrunning{V. Gaebel \textit{et al.}}

\maketitle

\section{Introduction}
\label{intro}
The exploration of the theory of strong interaction, called Quantum Chromodynamics (QCD), is one of the major goals of today's high energy physics. QCD is a non-abelian gauge theory that predicts a transition of the known \linebreak hadronic matter seen in nuclei at ground-state density to a fluid-like state called the Quark-Gluon-Plasma, QGP. This transition may either happen if a critical temperature around 150-160 MeV is reached or if a critical baryon density, around 4-5 times the ground state density, is created \cite{Harris:1996zx}. Naturally, this state appeared a few nanoseconds after the Big Bang and is currently present in compact stellar objects like Neutron Stars. Understanding the QGP and its features is one of the key tasks of today's research in nuclear physics. Heavy ion collisions are therefore carried out and investigated in particle accelerators and colliders at CERN, BNL, GSI and NICA.

The study of cluster formation processes in heavy ion collisions is of particular interest for a multitude of reasons: Firstly, clusters probe the two-particle baryon correlations in phase space, i.e. they allow to explore the space and momentum space structure of the emission source \cite{Monreal:1999mv}. Secondly, the production rate might allow to distinguish thermal production from coalescence \cite{Mrowczynski:2020ugu}. Thirdly, clusters, e.g. anti-matter, (multi-)strange objects like the hyper tritons, or even charmed clusters are themselves the objects of study and can be produced in hadronic collisions \cite{SchaffnerBielich:1999sy,Sun:2017unn,Braun-Munzinger:2018hat}.

As a first step to classify the production process, this paper investigates the formation of deuterons by the phase space coalescence of protons and neutrons. The phase space coalescence model has been shown to provide a good theoretical description of the formation of clusters in the considered energy ranges \cite{Mattiello:1995xg,Nagle:1996vp,Sombun:2018yqh,Sun:2018mqq,Sun:2020uoj}. The idea behind the coalescence model is that if a proton and a neutron are close enough in (momentum) space they can form a deuteron \cite{Schwarzschild:1963zz,Gutbrod:1988gt,Kapusta:1980zz,Scheibl:1998tk}.

In the coalescence picture, the probability of creating $N_{d}$ deuterons in a certain momentum space volume after freeze-out is proportional to the number of produced neutrons $N_{n}$ and protons $N_{p}$ (which can be further simplified, if one assumes the same number of protons and neutrons) \cite{Feckova:2016kjx}:

\begin{equation}
    N_d \sim N_n \cdot N_p \sim N_p^2
\end{equation}

Assuming, e.g. a thermal system and the same yield of protons and neutrons, we have $N_d \sim V$, $N_p \sim V$, and therefore $N_d/N_p^2 \sim B_2 \sim 1/V$. In this context, $B_2$ is called the coalescence parameter for deuteron production. I.e., the measurable quantity $B_{2}$ encodes information on the inaccessible spatial extent of the source. It is also clear that, in general, such a result might be modified by the details of the emitting source \cite{Mrowczynski:1993cx,Bleicher:1995dw}, correlations like flow and the internal wave function of the considered cluster \cite{Bazak:2018hgl}.

\section{The UrQMD model and coalescence}
For the theoretical investigation of the collisions, we perform simulations using the Ultrarelativistic Quantum \linebreak Molecular Dynamics (UrQMD) transport model in version 3.4 \cite{Bass:1998ca,Bleicher:1999xi}. This model has a well established history for the description of hadron yields and spectra over a broad range of energies (see e.g. \cite{Hillmann:2019wlt,Reichert:2019lny,Hillmann:2018nmd,Reichert:2020uxs,Hillmann:2019cfr}). 

\subsection{The model}
UrQMD is based on the covariant propagation of hadrons and their interactions by potentials and/or elastic and inelastic cross sections. UrQMD is either run in Boltzmann mode, i.e. it provides an effective solution to the relativistic Boltzmann equation or in the Hydro-Boltzmann hybrid mode (here abbreviated as ''UrQMD+hydro''). In hybrid mode, during the most dense phase of the reaction, the Boltzmann equation is replaced by an (ideal) fluid-dynamical evolution of the hot and dense QCD matter \cite{Rischke:1995ir,Rischke:1995mt,Petersen:2008dd}. In this mode, a phase transition to the QGP can be incorporated and compared to a purely hadronic treatment as well. 

The evolution equations in the hybrid mode are then

\begin{equation}
\partial_{\mu}T^{\mu\nu} = 0 ,
\end{equation}
\begin{equation}
\partial_{\mu}j_N^{\mu} = 0 ,
\end{equation}
with the energy-momentum tensor $T^{\mu\nu}$ and the baryon current $j_N^{\mu}$ \cite{Petersen:2008dd}. The initial state generated by UrQMD provides the equal time initial conditions for $T^{\mu\nu}$ and $j^{\mu}$. The hydrodynamic evolution is followed until the system reaches the freeze-out hyper-surface. There, we use a Cooper-Frye prescription to particlize \cite{Huovinen:2012is} the hydrodynamic cells stochastically. The propagation of the hadrons then proceeds in Boltzmann mode until kinetic freeze-out.

\subsection{Deuteron formation by coalescence}
For each event, UrQMD provides the 4-coordinates and 4-momenta of each hadron on the decoupling or freeze-out surface. Here freeze-out is defined for each hadron individually as the last space-time point of strong interaction, i.e. scattering or decay. A proton and a neutron are then assumed to form a deuteron if their distance in space and momentum space is sufficiently small. The details of the deuteron formation as implemented in UrQMD can be found in \cite{Sombun:2018yqh}. This method is used for both simulation modes, the Boltzmann mode and the hybrid mode with the intermediate hydrodynamic stage as described above. In the hybrid case, deuterons from the Cooper-Frye-hypersurface (direct thermal production) are not \linebreak taken into account, but only those formed later by coalescence in the late hadronic freeze-out stage \cite{Steinheimer:2012tb}.
\begin{figure}[t!]
\resizebox{0.5\textwidth}{!}{%
  \includegraphics{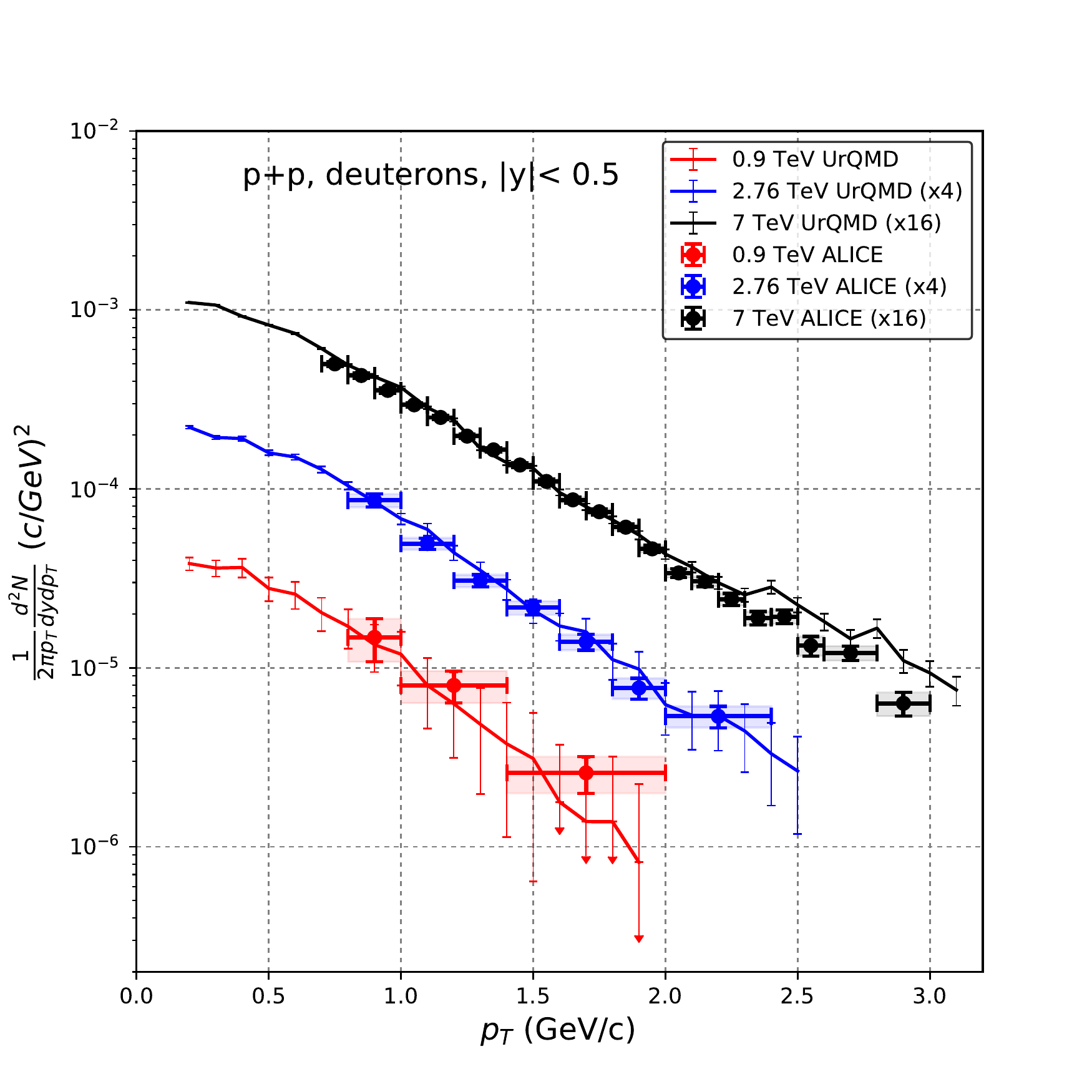}
}
	\caption[Deuteron yields]{[Color online] Invariant deuteron yields at midrapidity as a function of the transverse momentum $p_{T}$ in inelastic p+p collisions at different beam energies of $\sqrt{s_{NN}}$ = 0.9 TeV, 2.76 TeV and 7 TeV. The lines denote the UrQMD simulations and the circles the experimental data \cite{Acharya:2017fvb}.}
	\label{Fig1}
\end{figure}
\begin{figure}[t!]
\resizebox{0.5\textwidth}{!}{
	\includegraphics{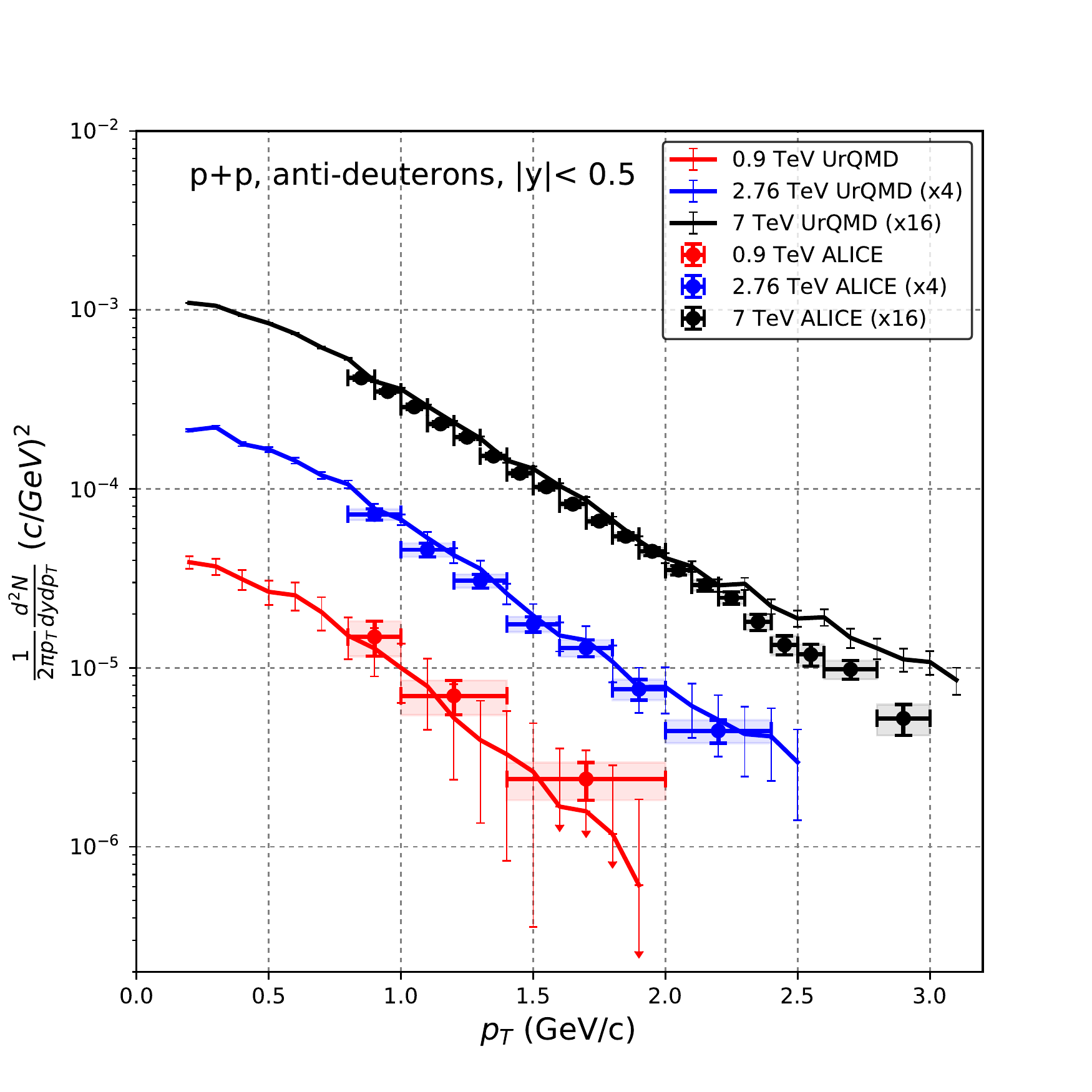}
	}
	\caption[Anti-Deuteron yields]
	{[Color online] Invariant anti-deuteron yields at midrapidity as a function of the transverse momentum $p_{T}$ in inelastic p+p collisions at different beam energies of $\sqrt{s_{NN}}$ = 0.9 TeV, 2.76 TeV and 7 TeV. The lines denote the UrQMD simulations and the circles the experimental data \cite{Acharya:2017fvb}.}
	\label{Fig2}
\end{figure}		
\subsection{Coalescence parameter $B_{2}$}
The coalescence parameter $B_2$ has been introduced in the simplified momentum space coalescence approach under the assumption that the volume can be trivially integrated out. In this approach one can connect the cluster distribution in momentum space to the distribution of the nucleons \cite{Scheibl:1998tk} via
\begin{equation}
	E_{A}\frac{dN_{A}}{d^3P_{A}} = B_{A} 
	\left(E_{p}\frac{dN_{p}}{d^3P_{p}}\right)^Z \left(E_{n}\frac{dN_{n}}{d^3P_{n}}\right)^N \bigg|_{P_{p}=P_{n}=P_{A}/A}.
\end{equation}
Here, $Z$ is the proton number, $N$ the number of neutrons and $A$ the mass number of the nucleus. $P_{A}$ is the momentum of the cluster and $P_{p} \ (P_{n})$ are the momenta of the protons (neutrons). Thus, the invariant momentum distribution of the cluster is proportional to the invariant momentum distributions of its constituents at the same momentum per particle. The coalescence factor is generally called $B_{A}$. 


This includes the assumption that the distribution is the same for protons and neutrons. Let us note that $B_{2}$ can be related (under certain assumptions) to the source size of the nucleons via \cite{Monreal:1999mv}
\begin{equation}
   R_{G}^{3} = \frac{3}{4} \pi^{\frac{3}{2}}\frac{m_{d}}{m_{p}^{2}}B_{2}^{-1},
\end{equation}
where $R_{G}$ denotes the Gaussian radius of the source.
Thus, the value of $B_2$ depends on the momentum distributions of the deuterons and the protons and indirectly encodes the spatial size of the proton source. 
\begin{figure}[t!]
\resizebox{0.5\textwidth}{!}{
	\includegraphics{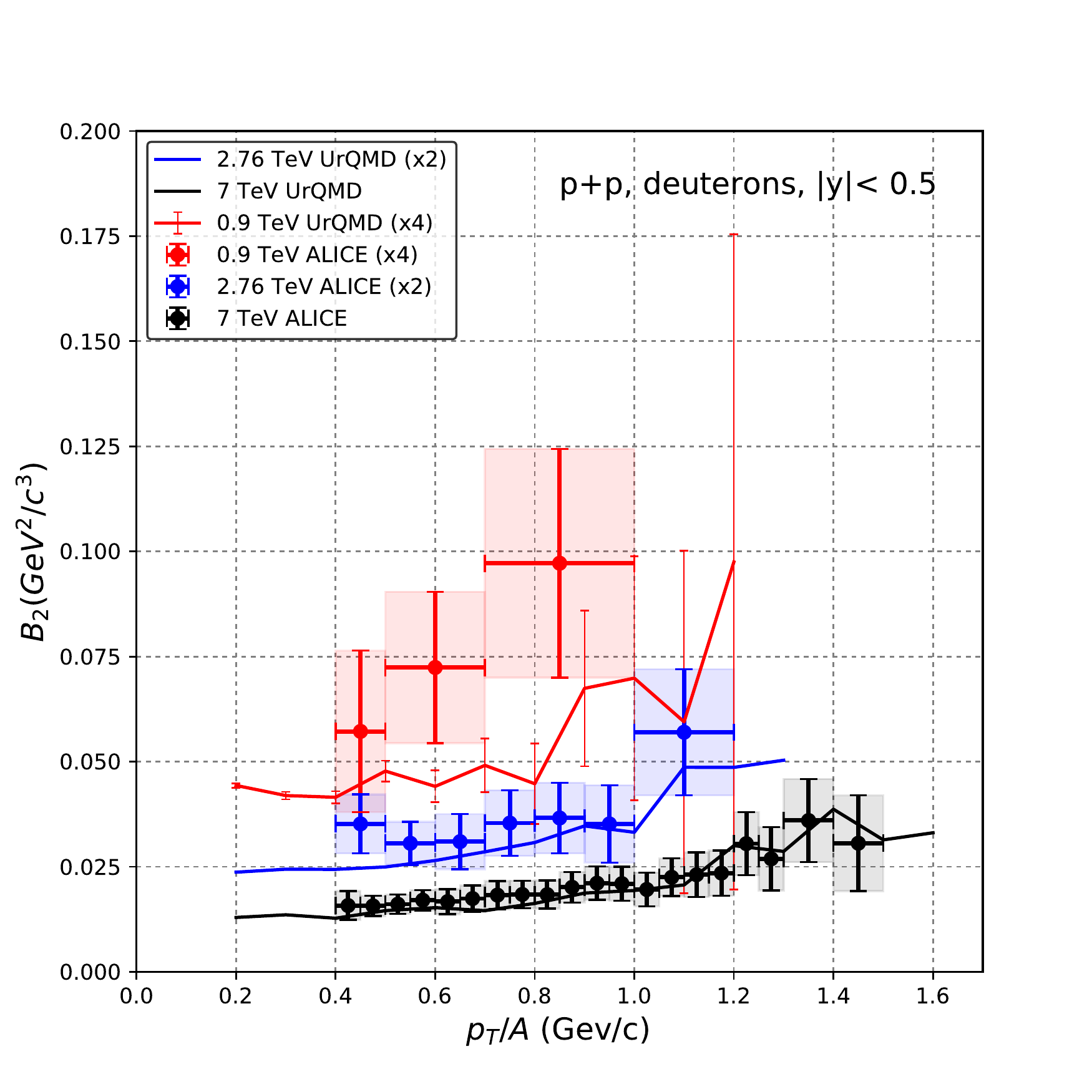}
	}
	\caption[$B_{2}$ against $p_{T}/A$]{[Color online] $B_{2}$ against the transverse momentum scaled with the mass number ($p_{T}/A$) in inelastic p+p collisions at beam energies $\sqrt{s_{NN}}$ = 0.9 TeV, 2.76 TeV and 7 TeV. The lines denote the UrQMD simulations and the circles the experimental data \cite{Acharya:2017fvb}.}
	\label{Fig3}
\end{figure}		
\begin{figure}[t!]
\resizebox{0.5\textwidth}{!}{
	\includegraphics{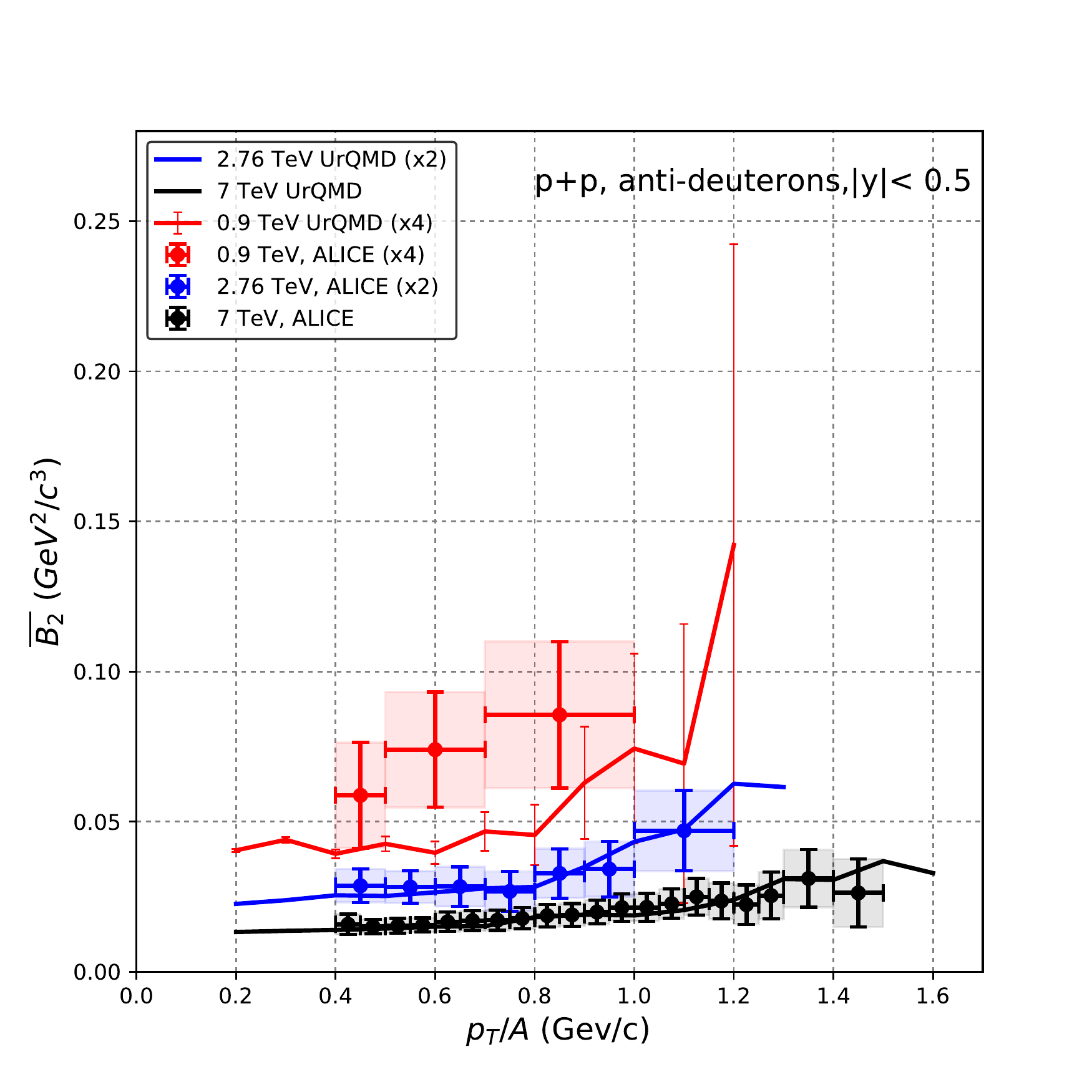}
	}
	\caption[ $B_{2}$ against $p_{T}/A$ ]{[Color online] $\bar{B_{2}}$ against the transverse momentum scaled with the mass number ($p_{T}/A$) in inelastic p+p collisions at beam energies $\sqrt{s_{NN}}$ = 0.9 TeV, 2.76 TeV and 7 TeV. The lines denote the UrQMD simulations and the circles the experimental data \cite{Acharya:2017fvb}.}
	\label{Fig4}
\end{figure}		

\section{Results}
 The following section presents comparisons of different experimental data (ALICE \cite{Acharya:2017fvb,Adam:2015vda,Acharya:2020sfy}, STAR \cite{Liu:2019nii,Adam:2019wnb}, E866 \cite{Adam:2019wnb}, E877 \cite{Adam:2019wnb} and PHENIX \cite{Adam:2019wnb}
)  to the simulation results from UrQMD. 

\subsection{Proton+Proton reactions}
Let us start with the exploration of proton+proton reactions in the TeV energy regime.
Fig. \ref{Fig1} shows the invariant yield of deuterons as a function of transverse momentum in inelastic proton+proton collisions for three different energies $\sqrt{s_{NN}}$ = 0.9 TeV, 2.76 TeV and 7 TeV (bottom to top). The lines indicate the UrQMD simulations and the symbols the experimental data \cite{Acharya:2017fvb}. Generally, we observe a good description of the experimental data in the investigated energy regime. Nevertheless, in collisions at $\sqrt{s_{NN}}$ = 7 TeV towards higher transverse momenta, the simulation tends to deviate slightly from the data. This might indicate a slight
overestimation of the jet cross section at high $p_{T}$.

Next, we turn to the invariant transverse momentum distribution of the anti-deuterons for p+p collisions at different beam energies of $\sqrt{s_{NN}}$ = 0.9 TeV, 2.76 TeV and 7 TeV (Fig. \ref{Fig2}).
The lines denote the UrQMD simulations and the circles the ALICE data points \cite{Acharya:2017fvb}. Again, we observe a good description of the data, with a slight deviation towards higher transverse momenta in $\sqrt{s_{NN}}$ = 7 TeV collisions. 

\begin{figure}[t!]	
\resizebox{0.5\textwidth}{!}{%
	\includegraphics{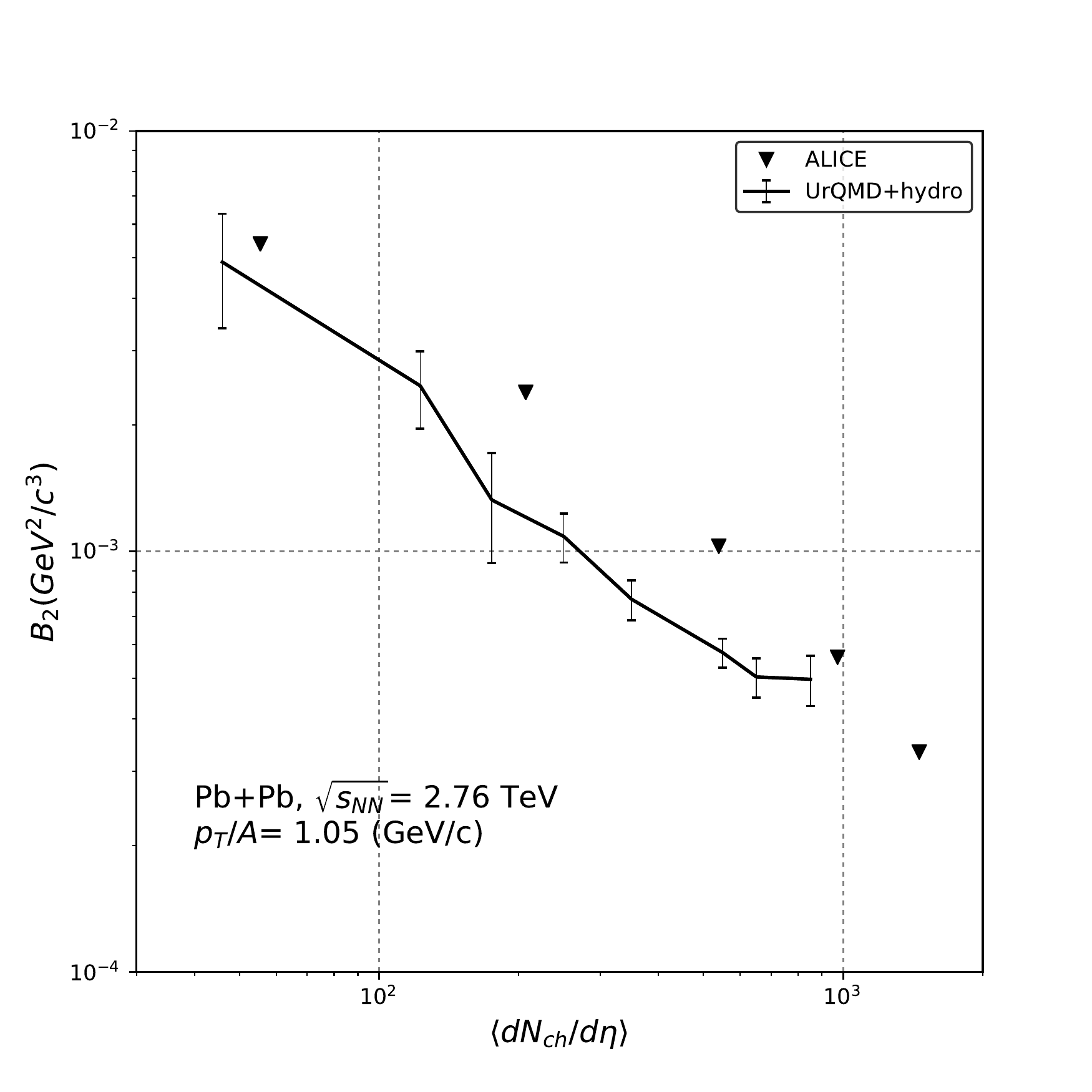}
	}
	\caption[$B_{2}$ against multiplicities at 2.76 TeV]
	{[Color online] $B_{2}$ as a function of the average charged-particle multiplicity in Pb+Pb collisions at fixed beam energy $\sqrt{s_{NN}}$ = 2.76 TeV at $p_{T}/A=1.05$ GeV/c. The black line denotes the UrQMD+hydro simulations and the black triangles the ALICE data points \cite{Acharya:2019rgc}.}
	\label{Fig5}
\end{figure}		

To allow for a better interpretation of the data, we explore in Fig. \ref{Fig3} and Fig. \ref{Fig4} the coalescence parameter $B_{2}$ for deuterons and $\bar{B_{2}}$ for anti-deuterons in inelastic proton+proton reactions at the same energies as above (i.e. $\sqrt{s_{NN}}=$ 0.9 TeV, 2.76 TeV and 7 TeV). Again, the lines denote the UrQMD simulations and the circles the experimental data \cite{Acharya:2017fvb}. In both cases, the data at $\sqrt{s_{NN}}$ = 0.9 TeV show a strong increase of $B_{2}$ and $\bar{B_{2}}$ with increasing $p_{T}/A$. However, for both deuterons and anti-deuterons this increase is only very mildly observed for the two higher energies. 
The present model calculations show a good description of $B_{2}$ and $\bar{B_{2}}$ as a function of the transverse momentum at $\sqrt{s_{NN}}$ = 2.76 TeV and $\sqrt{s_{NN}}$ = 7 TeV. Although the calculations at $\sqrt{s_{NN}}=$ 0.9 TeV increase with $p_{T}$, the simulated results are below the data points. 

Generally, (and in-line with standard HBT knowledge \cite{Monreal:1999mv,Akkelin:1995gh}) the effectively observed source size decreases towards higher transverse momenta. This volume effect in the data of proton+proton collisions at top LHC energies is also consistent with previous model calculations \cite{Scheibl:1998tk}.

\subsection{Nucleus+Nucleus reactions}
Let us next turn to nucleus+nucleus collisions. 
Here, the volume effect can be observed either in $B_2$'s centrality dependence at fixed energy or in the energy dependence at fixed centrality. 

The centrality dependence of $B_2$ is exemplified in Fig. \ref{Fig5} for Pb+Pb reactions at $\sqrt{s_{NN}}$ = 2.76 TeV. The black line denotes the UrQMD hybrid simulations and the black triangles the ALICE data points \cite{Schwarzschild:1963zz,Acharya:2020sfy}. Here, the charged particle density encodes the centrality, i.e. larger particle densities correspond to larger volumes. This is clearly reflected in the multiplicity dependence of $B_{2}$ which strongly decreases with increasing charge particle number. While the trend of the data is described well and supports the volume suppression of deuteron formation, the $B_{2}$ values in the simulations are slightly lower than in the ALICE experiment.
\begin{figure}[t!]	
\resizebox{0.5\textwidth}{!}{%
	\includegraphics{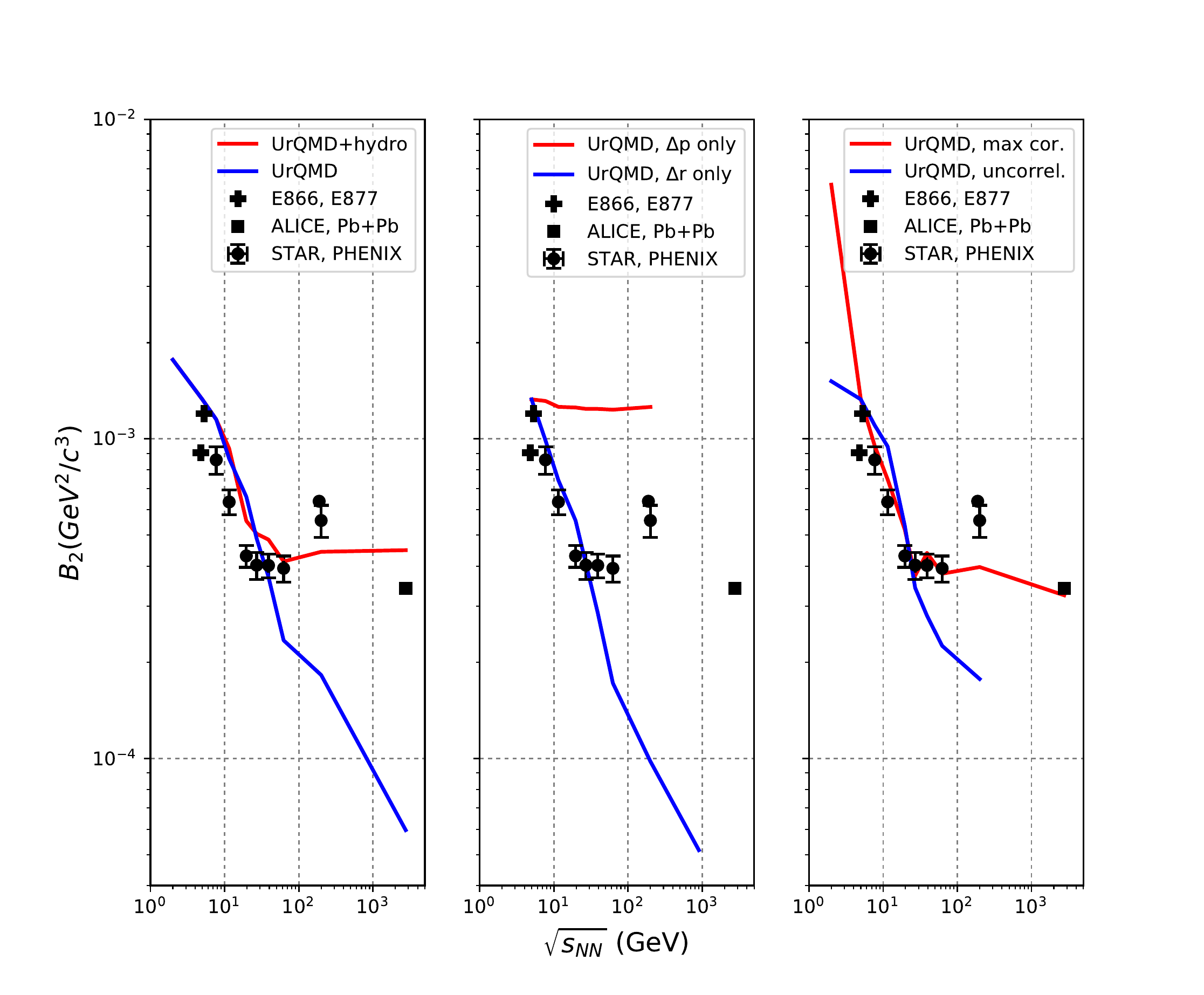}
	}
	\caption
	{[Color online] $B_{2}$ at midrapidity and at $p_{T}/A=0.65$ GeV/c as a function of the center-of-mass energy for Au+Au collisions. Additionally, the ALICE data point at $\sqrt{s_{NN}}$ = 2760 GeV for Pb+Pb collisions is shown. The black symbols denote the data from different experiments (E866 \cite{Adam:2019wnb}, E877 \cite{Adam:2019wnb}, STAR \cite{Adam:2019wnb}, PHENIX \cite{Adam:2019wnb}, ALICE\cite{Acharya:2019rgc}). Left panel: The red line shows the UrQMD+hydro simulations and the blue line the UrQMD cascade calculations. Middle panel: The red line indicates the UrQMD $B_{2}$ values for simulations using only momentum space coalescence, while the blue line shows the simulation results using only space coalescence. Right panel: The red line denotes the UrQMD simulations with maximal space-momentum correlations and the blue line the uncorrelated UrQMD calculations. }
	\label{Fig6}
\end{figure}		
In Fig. \ref{Fig6}, we contrast the centrality dependence of $B_{2}$ with the energy dependence for central collisions. Fig. \ref{Fig6} shows $B_{2}$ at midrapidity and at $p_{T}/A=$ 0.65 GeV/c as a function of the center-of-mass energy for Au+Au collisions. Additionally, the ALICE data point at $\sqrt{s_{NN}}$ = 2760 GeV for Pb+Pb collisions is shown. The black symbols denote data from different experiments \linebreak (E866 \cite{Adam:2019wnb}, E877 \cite{Adam:2019wnb}, STAR \cite{Adam:2019wnb}, PHENIX \cite{Adam:2019wnb}, ALICE \cite{Acharya:2019rgc}). The structure of the data can be summarized by a decrease of $B_{2}$ until $\sqrt{s_{NN}}$ = 20 GeV, followed by a constant $B_{2}$ value at high energies. We will now try to understand this structure by various analyses:

I) In the left panel of Fig. \ref{Fig6} we compare the data (symbols) to full UrQMD (blue line) and UrQMD+hydro (red line) simulations. For both calculations, we perform full phase space coalescence using $\triangle p_{max}$ = 0.285 GeV/c and $\triangle r_{max}$ = 3.575 fm. One clearly observes that \linebreak UrQMD+hydro with the full phase space coalescence approach is able to describe the data points nicely. Both regions, the strong decrease of $B_2$ and the constant level of $B_{2}$ above $\sqrt{s_{NN}}$ = 20 GeV are described. In contrast, the UrQMD simulation without hydrodynamics can only capture the decrease of $B_2$. 

II) We analyze this behavior in Fig. \ref{Fig6} (middle). Here we show the data (symbols) in comparison to calculations using either only momentum space coalescence (red line), i.e. $\triangle p_{max}$ = 0.285 GeV/c and $\triangle r_{max}$ = $\infty$ or only space coalescence (blue line) using $\triangle p_{max}$ = $\infty$ and $\triangle r_{max}$ = 3.575 fm (both curves are normalized to the theoretical $B_2$ value at $\sqrt{s_{NN}}$ = 5 GeV for better comparison). We clearly observe that the decrease until $\sqrt{s_{NN}}$ = 20 GeV is driven be the volume of the source \cite{Kittiratpattana:2020daw}. However, the flattening cannot be captured and the differences between the two UrQMD calculations cannot be explained.  

III) To pin down the origin of the flattening of the curve at higher energies, Fig. \ref{Fig6} finally compares the data (symbols) to UrQMD calculations with modified space-momentum correlations of the nucleons before coalescence. The idea is that for an expanding source the momentum $\vec p$ and the position $\vec r$ are correlated, if the transverse flow is sufficiently strong. For the correlation analysis, we compare a maximally space-momentum correlated nucleon source (constructed by enforcing  $\vec p || \vec r$) (red line) with a totally space-momentum uncorrelated source, constructed by randomly exchanging the momenta of the nucleons at different positions (shown as blue line). Again, both curves are normalized to the $B_2$ value at $\sqrt{s_{NN}}=$ 5 GeV for better comparison. One clearly observes that the uncorrelated nucleon source shows only a decrease, however, the correlated nucleon source shows the desired leveling-off and a plateau of $B_2$ with increasing energy.

Thus, we are led to the following conclusions: The volume effect dominates at low energies, leading to a decreasing $B_{2}$ value until $\sqrt{s_{NN}}$ = 20 GeV. Above $\sqrt{s_{NN}}$ = 20 GeV radial flow leads to substantial space-momentum correlations of the nucleons which result in a plateau of $B_2$ towards high energies. This also explains the difference between the UrQMD simulations with and without hydrodynamic stage, the main effect of the hydrodynamic stage is to produce sufficient flow to create the necessary space momentum correlations to capture the plateau structure of $B_2$.

\section{Summary}
In the present paper, we employed the UrQMD model to explore deuteron and anti-deuteron production in proton+proton and nucleus+nucleus reactions in the RHIC-BES and LHC energy regime. To this aim, the UrQMD model was supplemented with a phase space coalescence approach to form deuterons. The analysis has focused on the coalescence parameter $B_2$ that has an intuitive physical interpretation suggesting $B_2 \sim 1/V$. For proton+proton collisions, we observe an increase of $B_2$ with increasing transverse momentum indicating a smaller effective volume in line with our expectation from HBT correlations. In nucleus+nucleus collisions, we observe a) a strong centrality dependence of $B_2$ in line with expectations, and b) a strong energy dependence of $B_2$ at fixed centrality, which however levels off towards higher energies. We explain this non-monotonous structure as a volume effect up to $\sqrt{s_{NN}} \approx$ 20 GeV, counteracted  by strong space-momentum correlations at higher energies, which have been shown to create a plateau in $B_2$.

\section{Acknowledgments}
This work was supported by the Deutscher Akademischer Austauschdienst (DAAD), HIC for FAIR and in the framework of COST Action CA15213 THOR. The computational resources were provided by the Center for Scientific Computing (CSC) of the Goethe University Frankfurt.

%
\bibliographystyle{epj}
\bibliography{bibliography}

\end{document}